# Thermodynamic compatibility of actives encapsulated into PEG-PLA nanoparticles: *in silico* predictions and experimental verification

*Andreas Erlebach,[a,b] Timm Ott,[a] Christoph Otzen,[a] Stephanie Schubert,[b,c] Justyna Czaplewska,[b,d] Ulrich S. Schubert,[b,d] Marek Sierka[a,b]\**

[a] Otto Schott Institute of Materials Research (OSIM), Friedrich Schiller University Jena, Löbdergraben 32, 07743 Jena, Germany

[b] Jena Center for Soft Matter (JCSM), Friedrich Schiller University Jena, Philosophenweg 7, 07743 Jena, Germany

[c] Department of Pharmaceutical Technology, Institute of Pharmacy, Friedrich Schiller University Jena, Otto-Schott-Str. 41, 07745 Jena, Germany

[d] Laboratory of Organic and Macromolecular Chemistry (IOMC), Friedrich Schiller University Jena, Humboldtstrasse 10, 07743 Jena, Germany

KEYWORDS: PEG-PLA; polymeric nanoparticles; encapsulation efficiency; molecular dynamics; Flory-Huggins theory; atomistic simulations.

ABSTRACT

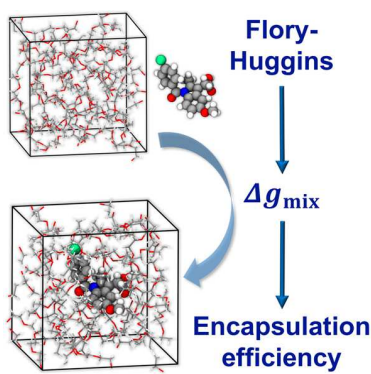

Achieving optimal solubility of active substances in polymeric carriers is of fundamental importance for a number of industrial applications, including targeted drug delivery within the growing field of nanomedicine. However, its experimental optimization using a trial-and-error approach is cumbersome and time-consuming. Here, an approach based on molecular dynamics simulations and the Flory-Huggins theory is proposed for rapid prediction of thermodynamic compatibility between active species and copolymers comprising hydrophilic and hydrophobic segments. In contrast to similar methods, our approach offers high computational efficiency by employing molecular dynamics simulations that avoid explicit consideration of the actual copolymer chains. The accuracy of the method is demonstrated for compatibility predictions between pyrene and nile red as model dyes as well as indomethacin as model drug and copolymers containing blocks of poly(ethylene glycol) and poly(lactic acid) in different ratios. The results of the simulations are directly verified by comparison with the observed encapsulation efficiency of nanoparticles prepared by nanoprecipitation.

**Introduction**

Solubility of low-molar-mass species in macromolecular compounds is of fundamental importance for a number of industrial applications, such as membrane separations, solvent extraction and thin film coating.[1] Within the growing field of nanomedicine, achieving optimal solubility of active substances (drugs) in polymeric nanoparticles represents a prerequisite for targeted delivery of anti-cancer therapeutics,[2] genes,[3] tumor imaging[4] agents and in theranostic applications combining simultaneously imaging and therapy.[3,5] An important parameter, which determines the encapsulation efficiency (EE) of nanoparticulate carriers, is the thermodynamic compatibility between a polymer and a drug. Its optimization by adjustment of polymer structures and compositions can lead to improved drug loading, retention and also the overall stability of the formulation. However, experimental optimization of EE using a trial-and-error approach is cumbersome and time-consuming. In contrast, atomistic computer simulations can not only efficiently predict the thermodynamic compatibility between polymers and drugs,[6] but also provide detailed understanding of the underlying interactions and structure-property relationships in polymer nanocomposites.[7] This knowledge can be employed as a powerful tool in the process of discovery and optimization of new drug delivery systems.[6,8-13]

The theoretical methods for predicting drug-polymer compatibility can be divided into analytical models and methods based on molecular dynamics (MD) simulations.[6,11-14] The first group includes various group contribution methods (GCM), which are based on an assumption that the total intermolecular interactions among molecules in a system can be represented as a linear sum of the contributions from various chemical groups within the molecules.[6] These methods are used for prediction of solubility parameters (SP), which provide an indication of the miscibility of two components. Although GCM-based SP are useful for rapid ranking of polymers based on their predicted relative ability to solubilize a particular drug, they do not account for entropic effects, concentration dependence and unique interactions between molecules.[6] Therefore, SP based on analytical models provide accurate predictions only for materials with similar structures.[15] Somewhat more accurate values of SP can be derived from MD simulations.[6,10,13,16]

A more reliable description of the compatibility between encapsulating polymer and a drug can be obtained within the Flory-Huggins (FH) theory, which comprises the mixing of two components as a process determined by enthalpic and entropic factors.[17-19] It includes a simplified description of (combinatorial) entropy contributions that allows the estimation of the free energy of mixing. The FH interaction parameter ($\chi_{FH}$), that enters the theory, is a dimensionless quantity, which represents the interactions contributing to the enthalpy of mixing of a polymer and a drug.[6] Two components are predicted to be miscible if $\chi_{FH} < 0.5$. The values of $\chi_{FH}$ can be obtained from SP calculated for the drug and polymer or using MD simulations.[6] In the former case the intrinsic errors of SP are carried over to $\chi_{FH}$. In contrast, calculation of $\chi_{FH}$ based on atomistic MD simulations allows for more reliable representation of the polymer structure and explicit inclusion of specific interactions between the drug and polymer chains.[6,12,13] Additionally, important effects such as hydrogen bonding, local arrangement of drug molecules within the micelle and differences in drug clustering around hydrophobic and hydrophilic polymer parts can be analyzed and quantified. This approach has been applied, e.g., to investigate the solubility of low molar mass substances in the dissolved state.[20] It has also been shown that the miscibility of different medical substances with block copolymers[10,11] and copolymer micelles[12] is well described by combining the FH theory with MD simulations, as demonstrated by a good agreement with experimental data.

For copolymers, computational studies of thermodynamic compatibility with active substances have been mainly limited to constant composition of the carrier material.[10-12,14,16] However, the selection of promising polymer candidates for specific drug carriers requires EE predictions for copolymers with varying composition, both in terms of hydrophilic and hydrophobic polymer types as well as different block lengths. In this study, we propose an approach based on MD simulations and the FH theory for the rapid prediction of the thermodynamic compatibility between active species and copolymers comprising hydrophilic and hydrophobic blocks. In contrast to similar methods,[10-12] our approach offers higher computational efficiency by employing MD simulations that avoid explicit consideration of the actual copolymer chains. The accuracy of the method is demonstrated for compatibility predictions between pyrene (PY) and nile red (NR) as model dyes as well as indomethacin (IMC) as model drug and copolymers containing blocks of poly(ethylene glycol) (PEG) and poly(lactic acid) (PLA) in different ratios. The results of the simulations are directly verified by comparison with observed EE of nanoparticles prepared by nanoprecipitation.

**Materials and Methods**

**Materials.** Reagents and solvents were purchased from Aldrich, Sigma and Fluka. 3,6-Dimethyl-1,4-dioxane-2,5-dione (lactide, LA) was recrystallized from ethyl acetate and dried under vacuum. Tin(II) 2-ethylhexanoate and dry methanol were used as received. Poly(ethylene glycol) methyl ether (2,000 (g/mol) and 5,000 (g/mol)) were dried under vacuum before reaction.

**Instrumentations.** $^1$H NMR spectra were measured on a Bruker 300 MHz using deuterated solvents. Size exclusion chromatography was performed using a SEC Agilent 1200 series, with a G1310A pump, a RID G1362A detector and a PSS GRAM guard/1000/30 Å (10 μm particle size) column at 40 °C. The system was calibrated with poly(ethylene glycol) methyl ether (PEG) standards. The eluent was dimethylacetamide (DMA) with addition of 0.21% of LiCl with a flow rate of 1 mL/min.

Dynamic light scattering was performed on a Zetasizer Nano ZS (Malvern Instruments, Herrenberg, Germany). Each measurement was performed in triplicate with each 3 × 30 runs at 25 °C ($\lambda$ = 633 nm, 173°). The mean particle size was approximated as the effective (Z-average) diameter and the width of the distribution as the polydispersity index of the particles (PDI) obtained by the cumulants method assuming a spherical shape of the nanoparticles.

The encapsulation efficiency (EE) of the actives was determined by UV/Vis spectroscopy (Infinite M200PRO, Tecan Group Ltd., Männedorf, Switzerland) of the centrifuged and lyophilized nanoparticle suspensions. The measurements were performed in DMSO, and the EE was quantified according to the emission of the pure active at $\lambda$ = 270 nm (pyrene), $\lambda$ = 555 nm (nile red), and $\lambda$ = 270 nm (indomethacin). All calibration functions show a linear correlation. The slight absorbance of DMSO itself at 270 nm did not affect the measurement.

**General procedure of synthesis of PLA.** The polymers (**PLA1**, **PLA2**) were synthesized according to literature in a slightly modified procedure.[21,22] 3,6-Dimethyl-1,4-dioxane-2,5-dione (**PLA1**: 7.88 g, **PLA2**: 15.8 g) was placed in the round bottomed flask and dried under vacuum for few hours. The catalyst (32 mg tin(II) 2-ethylhexanoate) and 12.5 μL of dry methanol were added under argon. The reaction mixture was heated until melted and stirred at 90 °C overnight. The product was dissolved in chloroform and precipitated into cold methanol.

**PLA1**: Yield: 6.7 g. $^1$H NMR (300 MHz, CDCl$_3$): $\delta$ (ppm) = 5.21-5.03 (CH, LA), 1.58-1.43 (CH$_3$, LA). SEC: $M_n$ = 16,190 g/mol; $M_w$ = 27,880 g/mol; PDI = 1.72.

**PLA2**: Yield: 13.4 g. $^1$H NMR (300 MHz, CDCl$_3$): $\delta$ (ppm) = 5.21-5.03 (CH, LA), 1.60-1.35 (CH$_3$, LA). SEC: $M_n$ = 39,990 g/mol; $M_w$ = 55,690 g/mol; PDI = 1.39.

**General procedure of synthesis of mPEG-PLA.** Copolymers were synthesized according to literature in a slightly modified procedure.[23-25] Poly(ethylene glycol) methyl ether was placed in the round bottomed flask and dried under vacuum for few hours at 40 °C. Subsequently, 3,6-dimethyl-1,4-dioxane-2,5-dione and the catalyst (tin(II) 2-ethylhexanoate) were added to the flask and dried further without heating. The reaction mixture was melted at 130 °C and stirred for further 7 h at 110 °C. The product was dissolved in chloroform and precipitated into cold methanol.

**PEG-PLA1:** 0.1 g mPEG (2,000 g/mol), 2.52 g 3,6-dimethyl-1,4-dioxane-2,5-dione, 10 mg tin(II) 2-ethylhexanoate. Yield: 1.13 g. $^1$H NMR (300 MHz, CDCl$_3$): $\delta$ (ppm) = 5.21-5.03 (CH, LA), 3.58 (PEG), 1.57-1.46 (CH$_3$, LA). $M_n$(NMR) = 35,700 g/mol. SEC: $M_n$ = 12,420 g/mol; $M_w$ = 24,460 g/mol; PDI = 1.97.

**PEG-PLA2**: 0.3 g mPEG (5,000 g/mol), 3.03 g 3,6-dimethyl-1,4-dioxane-2,5-dione, 12 mg tin(II) 2-ethylhexanoate. Yield: 2.45 g. $^1$H NMR (300 MHz, CDCl$_3$): $\delta$ (ppm) = 5.20-5.05 (CH, LA), 3.58 (PEG), 1.57-1.41 (CH$_3$, LA). $M_n$(NMR) = 53,240 g/mol. SEC: $M_n$ = 21,700 g/mol; $M_w$ = 44,650 g/mol; PDI = 2.06.

**PEG-PLA3**: 0.2 g mPEG (2,000 g/mol), 2.16 g 3,6-dimethyl-1,4-dioxane-2,5-dione, 20 mg tin(II) 2-ethylhexanoate. Yield: 1.47 g. $^1$H NMR (300 MHz, CDCl$_3$): $\delta$ (ppm) = 5.20-5.04 (CH, LA), 3.57 (PEG), 1.58-1.40 (CH$_3$, LA). $M_n$ (NMR) = 19,060 g/mol. SEC: $M_n$ = 7,920 g/mol; $M_w$ = 13,340 g/mol; PDI = 1.68.

**PEG-PLA4**: 0.5 g mPEG (5,000 g/mol), 2.16 g 3,6-dimethyl-1,4-dioxane-2,5-dione, 20 mg tin(II) 2-ethylhexanoate. Yield: 1.74 g. $^1$H NMR (300 MHz, CDCl$_3$): $\delta$ (ppm) = 5.20-5.04 (CH, LA), 3.57 (PEG), 1.59-1.39 (CH$_3$, LA). $M_n$ (NMR) = 24,370 g/mol. SEC: $M_n$ = 12,760 g/mol; $M_w$ = 20,920 g/mol; PDI = 1.63.

**Nanoparticle preparation.** For the encapsulation of actives (pyrene, nile red, indomethacin) into polymeric nanoparticles, 10 mg polymer and 0.2 mg active were dissolved in 2 mL THF. The solution was added dropwise to 5 mL water under continuous stirring. After removal of the THF by stirring for 2 days at room temperature, the nanoparticle suspensions were filtered using a 1 to 2 µm syringe filter in order to remove aggregated, not encapsulated actives.

**General Theory**

The FH theory employs a mean-field lattice approach that divides the mixed state into equally sized lattice sites (segments) representing subunits of active (**A**) and polymer (**P**). Each segment interacts with adjacent lattice sites with an average energy that depends on the binary interaction parameter $\chi_{\mathbf{A-P}}$ defined as

$$\chi_{\mathbf{A-P}} = \frac{\Delta h_{\text{mix}}}{RT\phi_{\mathbf{A}}\phi_{\mathbf{P}}}, \tag{1}$$

where $T$ is the temperature, $\phi_{\mathbf{A}}$ and $\phi_{\mathbf{P}}$ denote volume fractions of **A** and **P**, respectively, and $\Delta h_{\text{mix}}$ is the enthalpy of mixing. Neglecting volume change $\Delta h_{\text{mix}}$ can be approximated by the energy of mixing, $\Delta e_{\text{mix}}$,[26] which is defined as formation energy of the **A–P** state from the corresponding pure **A** and **P** components. Using the reference molar volume $v_r$ of one FH lattice site along with the volume fractions $\phi_{\mathbf{A}}$ and $\phi_{\mathbf{P}}$ of **A** and **P**, respectively, $\Delta e_{\text{mix}}$ can be calculated from cohesive energy densities $C_{\mathbf{A}}$, $C_{\mathbf{P}}$, and $C_{\mathbf{A-P}}$ of pure **A**, **P** and the mixture **A–P**, respectively, as

$$\Delta h_{\text{mix}} \approx \Delta e_{\text{mix}} = v_r(\phi_{\mathbf{A}} C_{\mathbf{A}} + \phi_{\mathbf{P}} C_{\mathbf{P}} - C_{\mathbf{A-P}}). \tag{2}$$

In case of copolymers, the key quantity for evaluation of thermodynamic compatibility with an active **A** is the (segment) molar Gibbs energy of mixing, $\Delta g_{\text{mix}}$. Here, we consider a copolymer **C** consisting of blocks of polymers $\mathbf{P}_1$ and $\mathbf{P}_2$. For a mixture of **C** and **A**, containing volume fractions $\phi_C = \phi_{P_1} + \phi_{P_2}$ and $\phi_A$ of **C** and **A**, respectively, such that $\phi_C + \phi_A = 1$, $\Delta g_{mix}$ is given as[27,28]

$$\Delta g_{mix} = RT \left( \frac{\phi_C}{r_C} \ln\phi_C + \phi_A \ln\phi_A + \phi_C \phi_A \chi_{eff} \right). \tag{3}$$

The parameter $r_C = r_{P_1} + r_{P_2}$ is obtained from ratios $r_{P_1}$ and $r_{P_2}$ of the molar volumes of **P**$_1$ and **P**$_2$ to **A**, respectively. Here, we assume that the volume $v_r$ of one segment (lattice site) refers to the molar volume of **A**. The effective interaction parameter $\chi_{eff}$ is a function of the volume fractions $f_{P_1}$ and $f_{P_1}$ of **P**$_1$ and **P**$_2$, respectively, that define the composition of **C** as well as the (binary) FH parameters $\chi_{AP_1}$, $\chi_{AP_2}$ and $\chi_{P_1 P_2}$

$$\chi_{eff} = f_{P_1} \chi_{AP_1} + f_{P_2} \chi_{AP_2} - f_{P_1} f_{P_2} \chi_{P_1 P_2}. \tag{4}$$

Approaches based on Eqs. (3) and (4) have been successfully applied for simulations of mixtures containing both random[29] and block copolymers.[30] $\chi_{eff}$ quantifies the interactions between **A** and **C** and can be used to calculate the Gibbs energy of mixing for different copolymer compositions assuming that the binary interaction parameters are independent of the polymer concentration. This way the thermodynamic compatibility between actives and copolymer nanoparticles whose hydrophobicity is controlled by their chemical composition can be calculated without performing explicit MD simulations of the actual copolymer chains. In contrast, previous studies investigated systems with fixed copolymer compositions using explicit MD simulations of the corresponding polymer chains.[10-12]

The values of cohesive energy densities required to calculate $\Delta e_{mix}$ in Eq. (2) can be derived from atomistic MD simulations. However, due to the large number of possible arrangements of an active and polymer chains efficient sampling of the most important configurations is of paramount importance. This is schematically depicted in Fig. 1, where a property $X$ such as cohesive energy or mass density is shown as a function of configuration space given by some generalized coordinates $\boldsymbol{q}$. $X$ as potential energy exhibits basins, or groups of shallow local minima, corresponding to different arrangements of an active and a polymer that are separated by higher energy barriers.[31] An MD simulation can sample only a limited number of local minima at a given temperature (shaded areas in Fig. 1). Two techniques can be employed to improve the efficiency of the sampling. The first one is to simply increase the size of the simulation cell and include several molecules of the active to sample different basins at once. The second approach employed in the present case is to use several independent MD simulations of smaller models that represent a particular subset of states within the configuration space. Time averages $x_i$ calculated for the corresponding MD trajectories are not ergodic with respect to the whole configuration space, *i.e.,* they do not correspond to the ensemble average of the system, but to that of the sampled basins (internal ergodicity).[31] Assuming Boltzmann statistics the approximation to ensemble average $\bar{X}$ can be calculated as a weighted sum of the time averages $x_i$ obtained from $n$ independent MD simulations

$$\bar{X} = \sum_{i=1}^{n} w_i\, x_i\,. \tag{5}$$

The Boltzmann weighting factors $w_i$ were calculated using the corresponding time averaged potential energies $E_i$

$$w_i = \frac{\exp(-\beta E_i)}{\sum_j \exp(-\beta E_j)}, \tag{6}$$

with $\beta = 1/RT$.

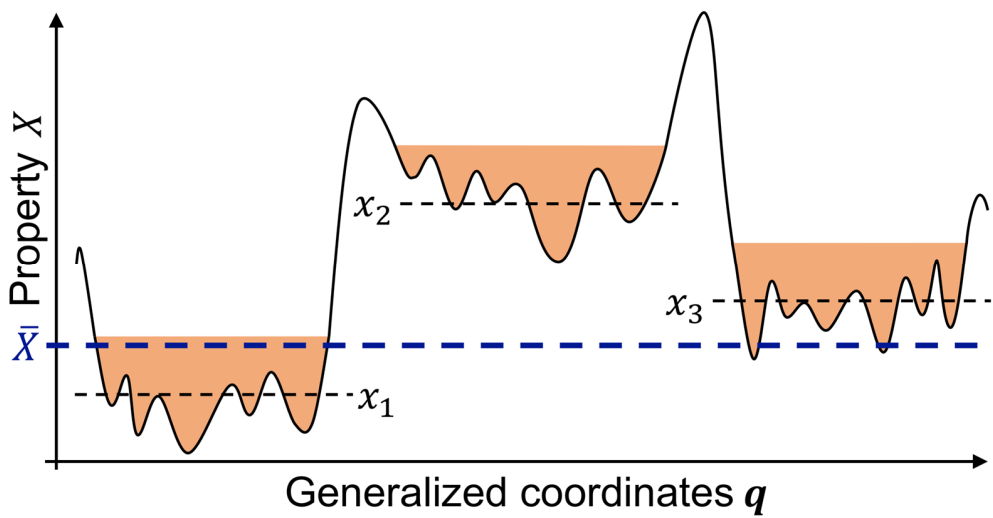

**Figure 1.** A property $X$ as a function of configuration space given by some generalized coordinates $\boldsymbol{q}$. $\bar{X}$ denotes Boltzmann weighted average. $X$ as potential energy exhibits basins, or groups of shallow local minima, corresponding to different arrangements of an active and polymer that are separated by higher energy barriers.

**Computational Procedure**

Figure 2 shows the computational procedure for calculation of the FH parameters and $\Delta g_{\text{mix}}$ that employs modules of the *Materials Studio*[32] (Version 8.0) programs suite along with the *ab initio* parameterized COMPASS force field.[33] The starting point are molecular models of **P** = **P**$_1$, **P**$_2$ and **A**. They are used for the construction of amorphous, three-dimensional unit cells of the mixed **A**–**P** states employing the algorithm of Theodorou and Suter[34] along with the Meirovitch scanning method implemented in the *Amorphous Cell* module.[35]

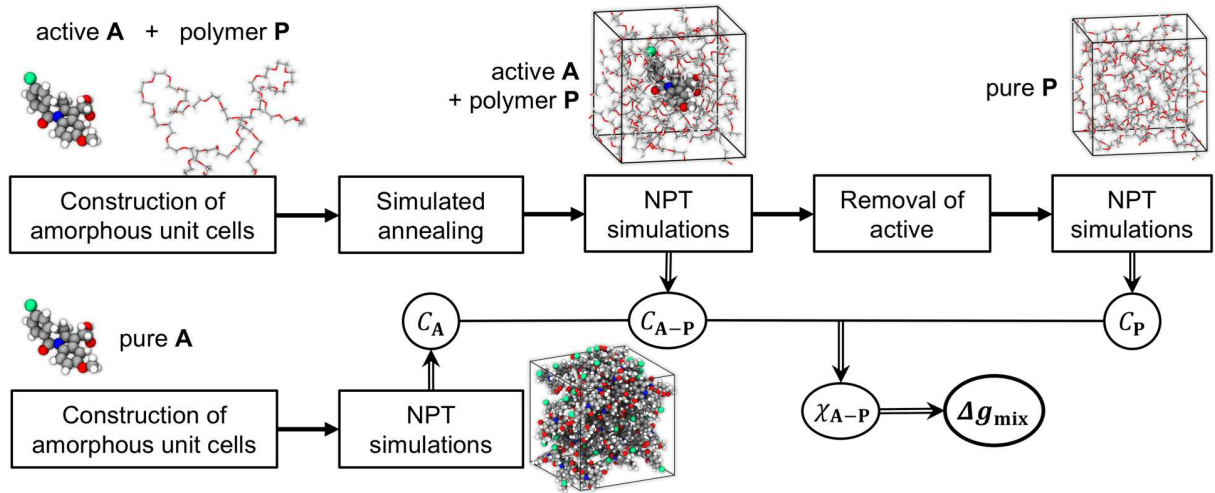

**Figure 2.** Computational workflow for the calculation of binary Flory-Huggins parameters, $\chi_{A-P}$, and the free energy of mixing, $\Delta g_{mix}$, for a copolymer comprising blocks of $P = P_1, P_2$ and an active **A** employing atomistic MD simulations. $C_A$, $C_P$, and $C_{A-P}$ denote cohesive energy densities of pure **A**, **P** and the mixture **A**–**P**, respectively.

In the next step, unit cells (15 in our case) with the lowest potential energies are selected for subsequent MD simulations. In order to generate energetically more stable configurations, an MD simulated annealing procedure is applied first employing the canonical (NVT) ensemble. For this, the models are equilibrated at $T$ = 298 K, than $T$ is raised in three steps to 500 K and later decreased in three steps back to 298 K. During each step, systems are allowed to evolve for 5 ps, with exception of the last step, where the equilibration time of 30 ps is applied. In order to allow changing of polymer conformations for generation of amorphous polymer unit cells the annealing temperature of 500 K exceeds the melting points of PEG (290-340 K)[36] and PLA (420-460 K).[37] The lowest energy structures (five in our case) obtained from the simulated annealing are then used to calculate the cohesive energy density $C_{A-P}$ of the **A**–**P** mixture employing MD simulations and the isothermal-isobaric (NPT) ensemble with a simulation time of 1 ns at $T$ =

298 K. In order to calculate the cohesive energy density $C_\mathbf{P}$ of the pure polymer **P**, **A** is removed from the system followed by an equilibration with the same setup as the preceded MD simulation. Finally, amorphous cells of pure **A** are constructed using an analogous approach and employed to calculate the corresponding cohesive energy density $C_\mathbf{A}$.

All MD simulations were performed with 1 fs time step employing the module *Forcite* along with the Nosé-Hoover thermostat.[38,39] NPT simulations used the Berendsen barostat[40] with zero target pressure. Time averages $x$ of appropriate properties were evaluated using the last 250 ps of each NPT simulation. The mean potential energy of the systems remained approximately constant over this period of time indicating the sufficient equilibration.

The calculation of $\chi_{\mathbf{P}_1\mathbf{P}_2}$ used amorphous unit cells containing PEG and PLA models with 150 and 5 repeating units, respectively. Since the molar volume of the PLA pentamer is similar to molar volumes of **A** that were used reference volumes $v_r$ in Eq. (2), its use avoids too drastic volume changes during NPT simulations after its removal from the system. Therefore, the PLA pentamer is an appropriate model representing one Flory-Huggins lattice site that approximates a segment of a PLA polymer chain neglecting the influence of end groups. For remaining calculations, PEG and PLA models containing 150 repeating units were employed. MD simulations of the active components (PY, NR and IMC) used for computation of thermodynamic compatibilities assume their amorphous (liquid like) state. This is consistent with the FH theory, which applies to random (amorphous) mixtures.[28] The amorphous reference state is also consistent with the amorphous structure of encapsulating nanoparticles prepared by nanoprecipitation.

**Results and Discussion**

**Computational results.** Table 1 summarizes the calculated mass densities $\rho$ and Hildebrand solubility parameters $\delta = \sqrt{C}$, where $C$ is the cohesive energy density, for pure PEG and PLA as well as PY, NR and IMC along with available experimental data. According to the theory of Hildebrand and Scott for regular solutions,[41,42] the FH parameter is directly connected with the solubility parameters $\delta_1$ and $\delta_2$ of the pure components and can be roughly approximated by using $RT\chi_{1-2} = v_r(\delta_1 - \delta_2)^2$. The table includes also results obtained for additional NPT simulations (1 ns at $T$ = 298 K) of crystalline PY[43] and NR.[44] In all cases the calculated values of $\rho$ and $\delta$ are in a very good agreement with experimental data supporting reliability of our simulation protocol.

**Table 1.** Comparison of calculated and experimental mass density $\rho$ (g/cm$^3$) and solubility parameter $\delta$ (MPa$^{1/2}$) for poly(ethylene glycol) (PEG), poly(lactic acid) (PLA), pyrene (PY), nile red (NR) and indomethacin (IMC).

|  | $\rho$ | | $\delta$ | |
| --- | --- | --- | --- | --- |
|  | Calc. | Exp.[a] | Calc. | Exp.[b] |
| PEG | 1.14 | 1.12 | 20.2 | 18.2 - 22.2 |
| PLA | 1.21 | 1.25 | 19.2 | 19.0 - 21.1 |
| PY | 1.17 (1.31)[c] | - (1.27)[c] | 22.5 | - |
| NR | 1.18 (1.29)[c] | - (1.39)[c] | 21.9 | - |
| IMC | 1.30 | 1.33 | 24.2 | - |

[a] PLA: ref. 37; PEG: ref. 45; PY: ref. 43; NR: ref. 44; IMC: ref. 46.

[b] PLA: ref. 47; PEG: ref. 36.

[c] Values refer to crystalline structures.

Furthermore, Figure 3 shows the convergence of the binary interaction parameters $\chi_{\text{PY-PEG}}$, $\chi_{\text{PY-PLA}}$ and $\chi_{\text{PEG-PLA}}$ with respect to the duration of the final NPT simulations using average quantities evaluated for the preceding 250 ps at each simulation time. The figure includes also the corresponding effective FH parameter $\chi_{\text{eff}}$, *cf.* Eq. (4), calculated for a PEG-PLA copolymer with $f_{\text{PEG}} = 0.1$. Starting from the simulation time of 1 ns, all interaction parameters are virtually constant showing the sufficient equilibration of the employed molecular dynamics simulations.

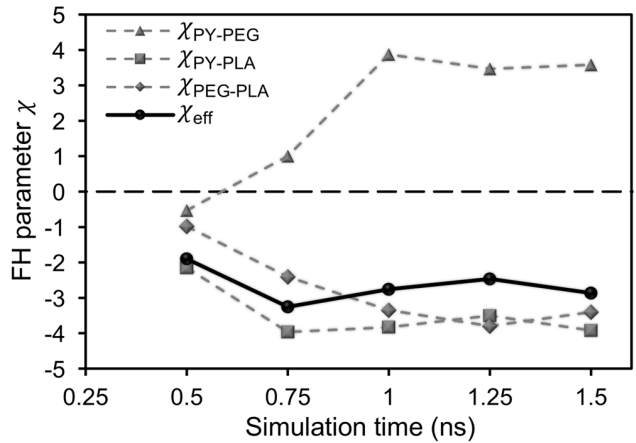

**Figure 3.** Convergence of binary interaction parameters and the corresponding effective FH parameter $\chi_{\text{eff}}$ (*cf.* Eq. 4) with respect to the duration of final NPT simulations for the mixture containing PY and an exemplary PEG-PLA copolymer with $f_{\text{PEG}} = 0.1$.

Table 2 shows the calculated FH parameters $\chi_{\text{A-P}}$ for each combination of active $\textbf{A}$ = PY, NR, IMC and pure polymer $\textbf{P}$ = PEG, PLA along with the corresponding interaction parameters $\chi_{\text{PEG-PLA}}$ for PEG–PLA mixture. The three different values of $\chi_{\text{PEG-PLA}}$ correspond to different molar volumes of $\textbf{A}$ taken as the segment reference volume $v_r$ in Eq. (2). Table 3 displays the calculated values of segmented molar Gibbs energy of mixing, $\Delta g_{\text{mix}}$, for the experimentally synthesized polymer samples and actives (PY, NR, IMC) along with the corresponding PEG volume fractions.

**Table 2.** Calculated binary FH interaction parameters $\chi$. Different values of $\chi_{\text{PEG-PLA}}$ correspond to different molar volumes of $\textbf{A}$ taken as the segment reference volume $v_r$ in Eq. (2).

| A | $\chi_{\text{A-PLA}}$ | $\chi_{\text{A-PEG}}$ | $\chi_{\text{PEG-PLA}}$ |
|---|---|---|---|
| PY | -3.83 | +3.87 | -3.35 |
| NR | +4.12 | +2.74 | -5.23 |
| IMC | -2.30 | -3.65 | -5.38 |

**Table 3.** Calculated segmented molar Gibbs energy of mixing $\Delta g_{\text{mix}}$ (J/(mol segment)) for each polymer sample and the corresponding PEG volume fractions ($f_{\text{PEG}}$).

| Polymer | $f_{\text{PEG}}$ | $\Delta g_{\text{mix}}$ | | |
|---|---|---|---|---|
| | | PY | NR | IMC |
| **PLA1** | 0.00 | -383 | 12 | -287 |
| **PLA2** | 0.00 | -383 | 13 | -286 |
| **PEG-PLA1** | 0.06 | -351 | 22 | -276 |
| **PEG-PLA2** | 0.10 | -330 | 28 | -270 |
| **PEG-PLA3** | 0.11 | -325 | 29 | -269 |
| **PEG-PLA4** | 0.21 | -274 | 39 | -257 |

**Experimental results.** PLA was prepared in two different lengths by ring-opening polymerization using a tin(II) catalyst. In addition, two PEG macroinitiators ($M_w$ = 2,000 g/mol for **PEG-PLA1** and **PEG-PLA3**; $M_w$ = 5,000 g/mol for **PEG-PLA2** and **PEG-PLA4**) were used for the synthesis of PEG-*b*-PLA copolymers in order to increase the hydrophilicity of the nanoparticle forming system. Consequently, two PLA samples and four PEG-PLA samples with varying hydrophobicity were prepared (Table 4). The polymers were all tested regarding their ability to encapsulate PY, NR and IMC into nanoparticles *via* nanoprecipitation. The polymers self-assemble in this process into spherical nanoparticles without the use of surfactants or other additives simply by dropping the diluted polymer solution into the non-solvent water while encapsulating other hydrophobic substances. In case of PEG-PLA, the PEG chains are mainly oriented to the surface of the nanoparticles. The formed PLA nanoparticles are in the range of 110 to 210 nm showing monomodal size distributions, whereas the PEG-PLA samples form smaller nanoparticles (70 to 120 nm). The initial concentration of actives in the nanoparticle formulation was 2 wt%. NR showed very low encapsulation efficiencies for the nanoparticle formation, and no relation to the hydrophobicity of the polymers itself can be found. PY and IMC were encapsulated in good yields with similar values with two exceptions. PY showed a significantly lower EE (25%) with **PEG-PLA4**, and IMC had a comparatively high EE with **PEG-PLA2**.

**Table 4.** Characterization of nanoparticle suspensions prepared by nanoprecipitation: polymer type, PEG volume fraction ($f_{PEG}$), copolymer molar mass ($M_w$, g/mol), nanoparticle size[b] (nm) and encapsulation efficiency[c] (EE, %).

| Polymer | $f_{PEG}$ | $M_w$[a] | PY Size (PDI) | EE | NR Size (PDI) | EE | IMC Size (PDI) | EE |
|---|---|---|---|---|---|---|---|---|
| **PLA1** | 0.00 | 28,000 | 130 (0.083) | 63 | 170 (0.074) | 3 | 110 (0.079) | 37 |
| **PLA2** | 0.00 | 56,000 | 150 (0.123) | 76 | 210 (0.071) | 3 | 170 (0.059) | 36 |
| **PEG-PLA1** | 0.06 | 24,000 | 90 (0.167) | 62 | 110 (0.098) | 3 | 90 (0.167) | 39 |
| **PEG-PLA2** | 0.10 | 45,000 | 120 (0.112) | 53 | 110 (0.135) | 3 | 120 (0.061) | 53 |
| **PEG-PLA3** | 0.11 | 13,000 | 90 (0.099) | 65 | 110 (0.106) | 2 | 80 (0.119) | 46 |
| **PEG-PLA4** | 0.21 | 21,000 | 70 (0.315) | 25 | 80 (0.159) | 3 | 70 (0.212) | 44 |

[a] Rounded values determined by SEC in Z-average diameter.

[b] Z-average diameter determined by DLS.

[c] Encapsulation efficiency determined by UV/Vis spectroscopy; EE of 100% is equivalent to 2 wt% of active in particle.

**Discussion**

In addition to the binary FH parameters summarized in Table 2, Figure 4 shows observed EE as function of segmented molar Gibbs energy, $\Delta g_{mix}$, and the experimental copolymer compositions, i.e. $f_{PEG}$ in Tables 3 and 4. The values of $\Delta g_{mix}$ are calculated for mixtures containing 2 wt% of **A** at $T = 298$ K using degrees of polymerization determined by the experimental molar masses of the PEG-PLA samples (Table 4).

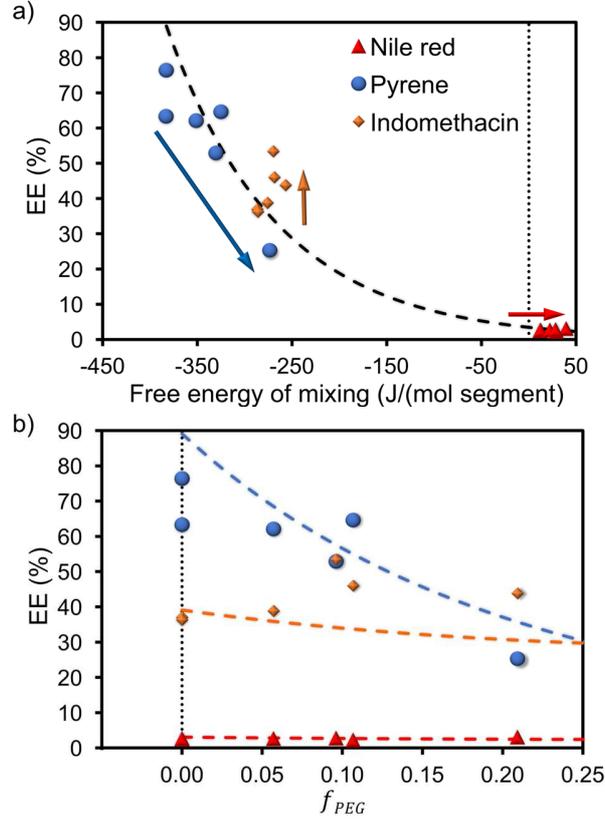

**Figure 4.** Experimental encapsulation efficiency (EE) as a function of: a) Gibbs energy mixing ($\Delta g_{\text{mix}}$) (arrows indicate the direction of increasing PEG fraction from 0.00 up to 0.21), and b) copolymer composition given as PEG volume fraction $f_{\text{PEG}}$. Dashed lines represent EE calculated by Eq. (8) with $c_1 = 3.417\%$, $c_2 = 8.116$ mol/kJ and $T = 298$ K.

An explicit relation between predicted $\Delta g_{\text{mix}}$ and observed EE can be provided by considering the transition of active **A** from its amorphous, liquid state $\boldsymbol{A}_{(l)}$ into the mixed (encapsulated) state $\boldsymbol{A}_{(enc)}$ according to the reaction $\boldsymbol{A}_{(l)} \rightleftharpoons \boldsymbol{A}_{(enc)}$ with the equilibrium constant $K_{eq}$:

$$K_{eq} = exp\left(-\frac{\Delta g_{mix}}{RT}\right). \tag{7}$$

Since the EE increases with larger values for $K_{eq}$, *i.e.* EE $\propto K_{eq}$, the dependency $EE = f(\Delta g_{mix})$ can be written as:

$$EE(\Delta g_{mix}) = c_1 exp\left(-\Delta g_{mix}\left(c_2 + \frac{1}{RT}\right)\right), \qquad (8)$$

with $c_1$ and $c_2$ as adjustable parameters. Least square fitting procedure yield for $c_1$ = 3.417% and $c_2$ = 8.116 mol/kJ at $T$ = 298 K along with a correlation coefficient of 0.974 showing the very good agreement of predicted and experimentally observed EE. The resulting curve for $EE(\Delta g_{mix})$ is shown in Figure 4a. In addition, this relation can be expressed as $EE(f_{PEG})$ by using Eqs. (3) and (4) for each mixture (Fig. 4b).

The negative values of the FH parameters $\chi_{PEG-PLA}$ (*cf.* Table 2) indicate miscibility of PEG and PLA at 298 K. Indeed, experiments have shown that amorphous PEG-PLA blends prepared by precipitation are at least partially miscible.[48] For NR both parameters $\chi_{NR-PLA}$ and $\chi_{NR-PEG}$ are strongly positive leading to $\Delta g_{mix} > 0$ and indicating the immiscibility of NR and PEG-PLA copolymers over the whole range of $f_{PEG}$. This is consistent with the very low EE determined experimentally (Table 4). In case of IMC both $\chi_{IMC-PLA}$ and $\chi_{IMC-PEG}$ are negative leading to $\Delta g_{mix} < 0$ for all PEG-PLA that shows virtually no variation with respect to $f_{PEG}$. The constant value of $\Delta g_{mix}$ is in agreement with the little dependence of EE on the copolymer composition observed in experiments. Furthermore, the lower value for $\chi_{IMC-PEG}$ rationalizes the slight increase of EE with growing PEG content in the nanoparticles. In contrast to NR and IMC, $\Delta g_{mix}$ for PY and PEG-PLA copolymers shows strong dependence on $f_{PEG}$ due to the opposite sign of the two FH parameters, $\chi_{PY-PLA}$ and $\chi_{PY-PEG}$. Therefore, pronounced dependence of EE on copolymer composition is expected since lower $\Delta g_{mix}$ should result in a higher thermodynamic preference to form a mixture of PY and PEG-PLA during precipitation of nanoparticles. This is indeed confirmed by our experimental data, as demonstrated in Fig. 4a and by data included in Table 4. The EE of PY in PEG-PLA changes from 25% for $f_{PEG} = 0.21$ to 76% for $f_{PEG} = 0.00$ (**PLA2**). This change is accompanied by an increase of $\Delta g_{mix}$ from -274 to -383 J/(mol segment). The hydrophobic compounds PY and NR show a considerably different solubility in PLA as indicated by binary interaction parameters $\chi_{PY-PLA}$ = -3.83 and $\chi_{NR-PLA}$ = +4.12, respectively. However, similar FH parameters were found for both dyes in combination with PEG. Such strong dependence of the solubility of PY and NR on the encapsulating polymer was also observed in experimental studies.[49]

These results demonstrate that our method can predict and quantify thermodynamic compatibility between active and copolymers with different chemical composition. The calculated values for $\Delta g_{mix}$ are a measure for the thermodynamic driving force to form polymer-active mixtures, *i.e.*, by assuming thermodynamic equilibrium large negative values for $\Delta g_{mix}$ should lead to an EE of 100% in the thermodynamic limit (large particles or bulk systems). However, due to smaller contribution of the copolymer to the total entropy of mixing (first term in Eq. (3)) compared to that of the active, the variation of the polymer chain length has less pronounced influence on $\Delta g_{mix}$. Therefore, pure PLA nanoparticle samples (**PLA1** and **PLA2**) containing PLA with different molar masses as well as **PEG-PLA2** and **PEG-PLA3** samples with almost equal $f_{PEG}$ show slightly different EE for PY despite nearly identical $\Delta g_{mix}$. In addition, the FH theory does not account for kinetic effects during the nanoparticle formation processes, which may be the main reason for the deviation between our theoretical predictions and experimentally determined EE. Improved description of thermodynamic compatibility can be obtained employing more sophisticated equation-of-state theories[50] instead of the relatively simple FH lattice model, and we intend to pursue further research in this direction. In addition, the sampling of configuration space can be improved whenever necessary by performing larger number of MD simulations or using larger simulation cells. However, the use of several smaller systems for MD simulations instead of a few large ones has the advantage of a trivial yet computationally efficient parallelization on multiple nodes of a computer cluster.

**Conclusions**

Experimental optimization of the loading capacity of polymeric nanocarriers using a trial-and-error approach is a cumbersome and time-consuming procedure. Instead, a combination of experiments and computer simulations can be employed as a powerful tool in the process of discovery and optimization of new polymeric delivery systems. Here, we have reported on an approach for rapid prediction of thermodynamic compatibility between active species and copolymers comprising hydrophilic and hydrophobic blocks. It is based on molecular dynamics simulations and the Flory-Huggins theory. In contrast to similar methods, our approach offers higher computational efficiency by employing molecular dynamics simulations that avoid explicit consideration of the actual copolymer chains. The accuracy of the method is demonstrated for compatibility predictions between pyrene and nile red as model dyes as well as indomethacin as model drug and copolymers containing blocks of poly(ethylene glycol) and poly(lactic acid) in different ratios. The results of the simulations are directly verified by comparison with observed encapsulation efficiency of nanoparticles prepared by nanoprecipitation. In agreement with the predicted compatibility, nile red is not encapsulated whereas indomethacin and pyrene show good solubility in the copolymer particles. Similarly, in line with theoretical predictions a pronounced dependence of the encapsulation efficiency of pyrene on the copolymer composition was found.

ASSOCIATED CONTENT

**Supporting Information**.

The following files are available free of charge.

AUTHOR INFORMATION

**Corresponding Author**

Prof. Dr. Marek Sierka, Otto Schott Institute of Materials Research (OSIM), Friedrich Schiller University Jena, Löbdergraben 32, 07743 Jena, Germany

**Author Contributions**

The manuscript was written through contributions of all authors. All authors have given approval to the final version of the manuscript.

**Funding Sources**

This work was carries out within the ProExcellence II initiative "NanoPolar" funded by the State of Thuringia, Germany.

ACKNOWLEDGMENT

This work was carried out within the ProExcellence II initiative "NanoPolar" funded by the State of Thuringia, Germany.

ABBREVIATIONS

PEG, poly(ethylene glycol); PLA, poly(lactic acid); PY, pyrene; NR, nile red; IMC, indomethacin; EE, encapsulation efficiency; MD, molecular dynamics; PDI, polydispersity index of the particles; SEC, size exclusion chromatography; NMR, nuclear magnetic resonance; FH, Flory-Huggins